\documentclass[journal]{vgtc}                % final (journal style)
\ifpdf%                                % if we use pdflatex
  \pdfoutput=1\relax                   % create PDFs from pdfLaTeX
  \usepackage{graphicx}                % allow us to embed graphics files
  \DeclareGraphicsExtensions{.pdf,.png,.jpg,.jpeg} % for pdflatex we expect .pdf, .png, or .jpg files
\else%                                 % else we use pure latex
  \usepackage{graphicx}                % allow us to embed graphics files
  \DeclareGraphicsExtensions{.eps}     % for pure latex we expect eps files
\fi%

\graphicspath{{figures/}{pictures/}{images/}{./}} % where to search for the images
\usepackage[table]{xcolor}
\usepackage{microtype}                 % use micro-typography (slightly more compact, better to read)
\PassOptionsToPackage{warn}{textcomp}  % to address font issues with \textrightarrow
\usepackage{textcomp}                  % use better special symbols
\usepackage{mathptmx}                  % use matching math font
\usepackage{times}                     % we use Times as the main font
\usepackage{cite}                      % needed to automatically sort the references
\usepackage{tabu}                      % only used for the table example
\usepackage{booktabs}                  % only used for the table example
\usepackage{tikz}

\usepackage{amsfonts}

\usepackage{enumitem}

\definecolor{cb_orange}{rgb}{1.0,0.51,0.0}
\definecolor{cb_blue}{rgb}{0.22,0.49,0.72}
\definecolor{cb_green}{rgb}{0.3,0.67,0.29}
\definecolor{cb_red}{rgb}{0.89,0.1,0.11}
\definecolor{cb_purple}{rgb}{0.6, 0.31, 0.64}
\definecolor{cb_brown}{rgb}{0.6, 0.4, 0.2}
\definecolor{cb_crimson}{rgb}{0.86, 0.08, 0.24}

\definecolor{our_blue}{HTML}{CADBEA}

\onlineid{0}

\vgtccategory{Research}
\vgtcpapertype{please specify}

\title{Visual Interaction with Deep Learning Models\\ through Collaborative Semantic Inference}

\author{Sebastian Gehrmann*, Hendrik Strobelt*, Robert Kr{\"u}ger,
Hanspeter Pfister, Alexander M. Rush}
\authorfooter{
\item 
(*) indicates equal contribution\\
\item
 S. Gehrmann and A. Rush are with the Harvard NLP group.
\item
 H. Strobelt is with IBM Research Cambridge
\item S.Gehrmann and H. Strobelt are with the MIT-IBM Watson AI Lab.
\item
 R. Kr{\"u}ger and H. Pfister are with the Harvard Visual Computing group.
}

\shortauthortitle{Gehrmann \MakeLowercase{\textit{et al.}}: Visual Interactions with Deep Models through Collaborative Semantic Inference}
\abstract{Automation of tasks can have critical consequences when humans lose agency over decision processes. Deep learning models are particularly susceptible since current black-box approaches lack explainable reasoning. We argue that both the visual interface and model structure of deep learning systems need to take into account interaction design. We propose a framework of collaborative semantic inference (CSI) for the co-design of interactions and models to enable visual collaboration between humans and algorithms. The approach exposes the intermediate reasoning process of models which allows semantic interactions with the visual metaphors of a problem, which means that a user can both understand and control parts of the model reasoning process. We demonstrate the feasibility of CSI with a co-designed case study of a document summarization system. %The study demonstrates how humans can retain agency without losing the benefits of using deep learning systems. 
} % end of abstract

\CCScatlist{ % not used in journal version
 \CCScat{K.6.1}{Management of Computing and Information Systems}%
{Project and People Management}{Life Cycle};
 \CCScat{K.7.m}{The Computing Profession}{Miscellaneous}{Ethics}
}

\renewcommand{\manuscriptnotetxt}{}

\vgtcinsertpkg

\begin{document}


\firstsection{Introduction}

\maketitle

%% \section{Introduction} %for journal use above \firstsection{..} instead
Deep learning has become a universal tool that is increasingly applied in automated approaches to commonplace problems. 
Often, improvements in performance and efficiency from deep models come at the cost of increased model complexity, which leads to difficulties in interpretation, analysis, and visualization. 
By relying on the outputs of these complex black-box models, users give up their agency and control over the automated process. 
Moreover, the users are forced to trust and rely on models that have been shown to be biased, and that can make inexplicable mistakes.\looseness=-1

% This is too cute
%Recent examples how a loss of agency can lead to harmful effects are the crashes of multiple aircraft that have been partially attributed to the increasing automation of the cockpit~\cite{wise_2019,potocnik2018tragic}. 

To combine the advantages of models and humans, researchers across disciplines have advocated for models as collaborative team members. Grosz argues that the development of ``intelligent, problem-solving partners is an important goal in the science of AI''~\cite{grosz1996collaborative}. Horvitz' principles for mixed-initiative user interfaces~\cite{horvitz1999principles} and the more recent guidelines for human-AI interaction~\cite{amershi2019guidelines} call for ``mechanisms for efficient agent-user collaboration to refine results.'' 
Similarly, Heer points out that work on automation ignores the design space in which computational assistance augments and enriches people's intellectual work~\cite{Heer1844}.

We propose a framework that can be applied to enable users to control predictive processes called \emph{collaborative semantic inference} (CSI). CSI describes a dialogue, alternating between model predictions presented in a visual form and user feedback on internal model reasoning. 
This process requires exposing the model's internal process in a way that mirrors the \textit{users mental model} of the problem and then empowering the user to influence this process.
This approach is centered around the core design principles for visual analytics, which integrates visualization and analytics in a human-centered interface~\cite{keim2008visual}. 
Endert et al.~\cite{endert2012semantic} define semantic interactions as those which  ``enable analysts to spatially interact with [such] models directly within the visual metaphor using interactions that derive from their analytic process.''
CSI describes \textit{how} to connect these semantic interactions to the model inference process.
The development of CSI methods and interfaces further requires a tight collaboration between the visual analytics, interaction design, and machine learning experts, which is a challenging, but promising, direction of research~\cite{sacha2019vis4ml,endert2017state,stolper2014progressive,hohman2018visual}.

Most deep neural networks do not expose their internal reasoning process. The CSI framework requires the development of model extensions that expose intermediate reasoning that can be associated with user-understandable choices. In this work, we present a proposal that incorporates discrete latent variables~\cite{kim2018tutorial} into the model design. % EXAMPLE HERE? 
%\rk{short example would be helpful, otherwise difficult to understand at this point in the paper.}
These variables act as ``hooks'' that can control the reasoning process and output of a model. The hooks enable what-if analyses by answering what internal choices would have led to a specific output. Crucially for CSI, the hooks also allow a user to infer the model's reasoning process by seeing how a given output was selected. This visual analysis in the backward direction, from prediction to input, is typically not possible without model modifications.

To contextualize the multi-disciplinary co-design process and assist in developing collaborative tools, we provide an overview of the visualization and interaction design space for neural network models.
We describe the actions required to move towards the goal of retaining human agency through visual CSI interfaces. We further connect our categorization to the previously described user roles of architect, trainer, and end-user~\cite{strobelt2018lstmvis} and a classification into interpretability methods that aim to understand and shape the model or its decisions. 
%Our categorization classifies approaches as passive, or interactive observation approaches in the absence of feedback mechanisms that link back to the model and we describe methods to enable collaborative interactions for observation tools.

As proof of concept, we apply our design process to the use case of a document summarization system. When this task is handled by an automated system, the results almost always require heavy post-editing by humans. Our use-case presents a first attempt at a collaborative, deep learning-based, interface for this problem. This use-case also demonstrates the expanded design space of interactive visual interfaces for collaborative models. We further identify challenges encountered in these collaborative models: How do we develop an interface that visualizes the prediction process on a granular level for any kind of model-specific input type (e.g.,\ text, images, spectrograms), and how can a visual interface provide an integration of human interventions as part of the prediction process.

The paper contains the following contributions. In Section~\ref{sec:design}, we define the \textbf{concept of collaborative semantic inference} and its place in the design space of integrating deep neural models with visually interactive interfaces. We then describe \textbf{how CSI can be incorporated} into models using latent variables in Section~\ref{sec:latvar}. We \textbf{showcase a visually interactive use-case} for applying CSI to a text summarization task in Section~\ref{sec:uc2}. Based on our experience gained from this use case, we describe learned lessons towards building a systematic co-design process (Section~\ref{sec:process}). We discuss the implications of CSI and its advantages and disadvantages in Section~\ref{sec:conclusion}.

\section{Design Space for Integrating a Machine-Learned Model and Interactive Visualization}
\label{sec:design}

While previous work has categorized and described the vast design space for visual interpretation of machine learning models~\cite{endert2017state, liu2017towards, lu2017state, hohman2018visual}, the problem of co-designing models, visualizations, and interactions remains challenging. 
In an analysis, Crouser and Chang find that there exists no common language to describe human-computer collaboration interfaces in visual analytics and propose to reason about interfaces based on the possible interactions humans can have with a model~\cite{crouser2012affordance}.  

Expanding this idea, we contextualize different co-design approaches by categorizing the design space based on three criteria: (1) the level of integration of a machine learning model and a visually interactive interface, (2) the user type, and (3) applications that aim to understand and shape the model or model-decisions. Table~\ref{tab:cate} shows a categorization of the related work using criteria (1) and (3).

\begin{figure}[t!]
    \centering
    \includegraphics{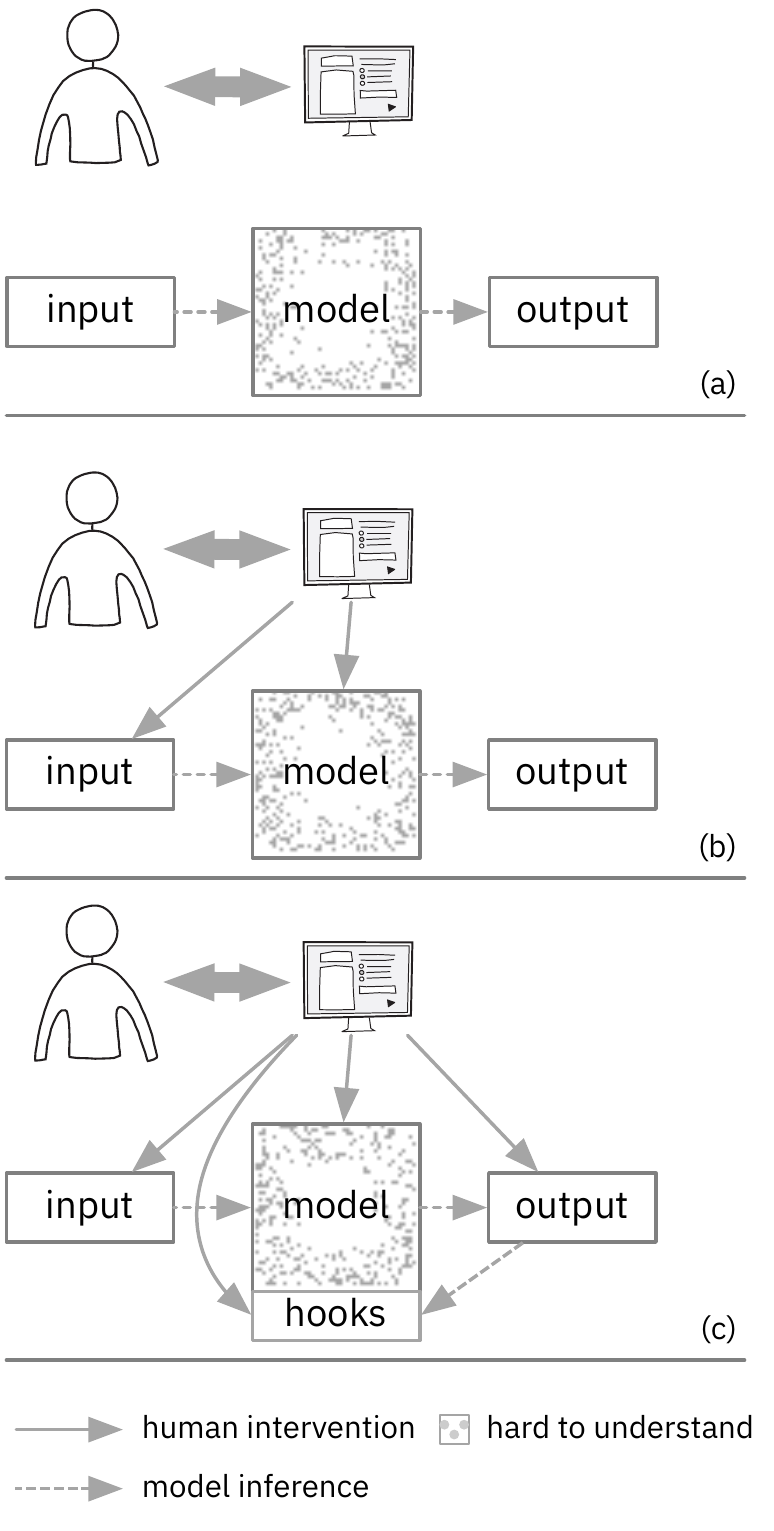}
    \caption{We define the three levels of integration between models and visual interfaces: (a) passive observation, (b) interactive observation, and (c) interactive collaboration. Each successive stage extends the design space by adding new potential interactions. In the interactive collaboration stage, the model itself is extended with ``hooks'' to enable semantic interactions.}
    \label{fig:intmodel}
\end{figure}

We identify three broad integration approaches between models and visual interfaces (Figure~\ref{fig:intmodel}): \emph{passive observations}, \emph{interactive observations}, and \emph{interactive collaborations}. 
As we will discuss in detail in this section, each category comprises a class of techniques that address different challenges of an analysis pipeline.
%The framework differentiates visualization techniques based on the type of possible interactions with a model, specifically to what extent a user can interact with the model and the model with the user. 

To decide on the visualizations and interactions for a specific problem in this design space we need to additionally consider the user type. We follow Strobelt et al.~\cite{strobelt2018lstmvis} who describe the roles of model \emph{architects}, \emph{trainers}, and \emph{end-users}. Architects are defined as users who are developing new machine learning methodologies or who adapt existing deep architectures for new domains. Trainers are those users who apply known architectures to tasks for which they are domain experts. 
End-users are those users who use trained models for various tasks and who may not know how a model functions. As domain experts, the end-users' main goal is to achieve and explain the results of a model.

% SG TODO: MAKE SUPER CLEAR THAT THIS IS ABOUT GOALS NOT CHOICES WHAT INTERACTIONS TO INCLUDE!
% Model-Understanding/Shaping: Systems with the goal of understanding or shaping the model through its features and parameters
%Decisions-Understanding/Shaping: Systems with the goal of understanding or shaping the individual decisions of the model on specific instances

In addition to the level of integration and user type, we divide the design space between visual interfaces to \emph{understand and shape the model or the decisions} of the model. 
Model-understanding/shaping describes systems with the goal of understanding or shaping the model through its features and parameters. The methods can help to gain insights into or modify the parameters that a model is in the process of learning or has learned already.
Decision-understanding/shaping systems have the goal of understanding or shaping the individual decisions of the model on specific instances. Decision-based applications aim to understand how the model arrives at a given output but does not modify the parameter of the model.
Shaping the model and decisions requires a tight coupling between the model and the interface, which in our classification is only possible with interactive collaboration interfaces, whereas the other categories focus on understanding. 

\subsection{Passive Observation}

The first stage in our design space is passive observation, in which visualizations present a user with static information about the model or its output. The information can target any user-type and range from global information about model training to heatmaps overlaid over specific inputs. 
Passive observation interfaces only require a loose coupling between the interface and the model (Figure~\ref{fig:intmodel}(a)).

\paragraph{Model-Understanding}

Architects and trainers are often concerned about how well a model is training, i.e., the model \emph{performance}. 
Tools can assist trainers by tracking and comparing different models and metrics, e.g., Tensorboard~\cite{wongsuphasawat2018visualizing}.
Moreover, it is crucial for trainers to understand whether a model learns a good representation of the data as a secondary effect of the training, and to detect potential biases or origins of errors in a model~\cite{belinkov2019analysis}.
To address this issue, many model-understanding techniques aim to visualize or analyze learned global features of a model~\cite{zeiler2014visualizing,carter2019activation, odena2016deconvolution, belinkov2017neural}.
% An associated measure of evaluating a tool for this outcome is whether the monitoring leads do easier decisions about restarting or aborting training runs or whether the visualizations lead to a better mental model of the model performance.

\begin{table*}[t!]
\centering
\begin{tabular}{@{} >{\columncolor{white}[0pt][\tabcolsep]}p{.2\textwidth} %p{.16\textwidth} 
p{.35\textwidth} >{\columncolor{white}[\tabcolsep][0pt]}p{.35\textwidth} @{}}
\toprule
 & %Interaction with\dots& 
 Model-Understanding/Shaping & Decision-Understanding/Shaping  \\ \midrule
Passive Observation\newline
\textcolor{gray}{[Understanding]}& %Input & 
Activation Atlas~\cite{carter2019activation}, 
Classification Visualization~\cite{chae2017visualization},
DeepEyes~\cite{pezzotti2018deepeyes},
Feature classifiers~\cite{belinkov2017neural},
Tensorboard~\cite{wongsuphasawat2018visualizing},
Weight Visualization~\cite{tzeng2005opening} & 
Deconvolution~\cite{zeiler2014visualizing}, 
LIME~\cite{ribeiro2016should}, 
Neuron Analysis~\cite{dalvi2019one}, 
Rationals~\cite{lei2016rationalizing}, 
Saliency~\cite{li2015visualizing}, 
Structured Interpretation~\cite{dey2015predicting},
Prediction Difference~\cite{DBLP:conf/iclr/ZintgrafCAW17}\\ \addlinespace[0.2cm]
Interactive Observation\newline
\textcolor{gray}{[Understanding]} & %Input, Model Internals & 
ActiVis~\cite{kahng2018acti},
Deep Visualization~\cite{yosinski2015understanding}, 
Embedding Projector~\cite{smilkov2016embedding}, 
GANViz~\cite{wang2018ganviz}, 
LSTMVis~\cite{strobelt2018lstmvis}, 
Prospector~\cite{krause2016interacting}, 
RetainVis~\cite{kwon2019retainvis}, 
RNNVis~\cite{ming2017understanding}, 
ShapeShop~\cite{hohman2017shapeshop} & 
Instance-Level Explanations~\cite{krause2017workflow},
Manifold~\cite{zhang2019manifold}, 
NLIZE~\cite{liu2019nlize},
RNNBow~\cite{cashman2017rnnbow},
Semantic Image Synthesis~\cite{chen2017photographic}, 
Seq2Seq-Vis~\cite{strobelt2019s}\\
\addlinespace[0.2cm]
Interactive Collaboration\newline
\textcolor{gray}{[Shaping]} & %Input, Model Internals, Hooks, Output &
Human-in-the-Loop Training~\cite{holzinger2016interactive}, 
Statistical Active Learning~\cite{cohn1996active},
Tensorflow Playground~\cite{smilkov2017direct},
Explanatory Debugging~\cite{kulesza2010explanatory}
&
%\cellcolor{our_blue}
\textbf{Achievable via CSI}, for example in GANPaint~\cite{bau2018gandissect} \\ \bottomrule
\end{tabular}%
\vspace{1em}
\caption{A classification of some of the related work, which shows the distinct lack of collaborative interfaces for decision-understanding and shaping. Our CSI framework aims to fill this void, with a particular focus on applications built for end-users. However, different user types span all of the previous work, some aiming towards architects, trainers, end-users, or a combination of them.}
\label{tab:cate}
\end{table*}

\paragraph{Decision-Understanding}

Decision-understanding passive observation tools assist end-users in developing a mental model of the machine learning model behavior for particular examples. The most commonly applied decision-understanding techniques present overlays over images~\cite{ribeiro2016should,dey2015predicting} or texts~\cite{dalvi2019one,lei2016rationalizing,li2015visualizing}. These overlays often represent the activation levels of a neural network for a specific input. For example, in image captioning, this method can show a heatmap that indicates which part of an image was relevant to generate a specific word~\cite{DBLP:conf/icml/XuBKCCSZB15}. 
%It is challenging to define measures that accurately reflect whether the techniques achieve the desired outcome of developing a mental model of the behavior. 
A qualitative assessment of these methods may focus on whether highlights match human intuition before and after changes to the input or model~\cite{2017arXiv171100867K,SanityNIPS2018}. These methods also commonly assist in a pedagogical context~\cite{hohman2018visual,smilkov2017direct}. 

\subsection{Interactive Observation}

Interactive observation interfaces can receive feedback or information from the model itself (Figure~\ref{fig:intmodel}(b)). This feedback enables the testing of multiple hypotheses about the model behavior within a single interface.
We classify tools as interactive observation that allow changing inputs or extracting any model-internal representation as part of an interactive analysis. We call these \emph{forward} interactions, analogous to how sending inputs into a model is called the forward pass. This approach can be used by trainers with domain knowledge to verify that a model has learned known structures in the data, or by end-users to gain novel insights into a problem.
The development of interactive observation methods has been an active field of research, as summarized in recent review papers~\cite{liu2017towards, lu2017state, hohman2018visual}.
Interactive observation allows for a richer space of potential interactions than passive observation tools and thus require a closer coupling between visualizations, interface, and the model~\cite{endert2017state}.

\paragraph{Model-Understanding}

In an extension of visualization of learned features, interactive observational tools enable end-users and trainers to test hypotheses about global patterns that a model may have learned. One example is Prospector~\cite{krause2016interacting}, which can be used to investigate the importance of features and learned patterns.  Alternatively, counterfactual explanations can be used to investigate changes in the outcome of a model for different inputs, thereby increasing trust and interpretability~\cite{whatif,wachter2017counterfactual}.
%The visual approaches to integrate an expert's domain knowledge range from user-guided instance and feature selection.

\paragraph{Decision-Understanding}

Interactive decision-understanding tools visualize how minor changes to an input or the internal representation influences the model prediction. 
Interactively building this intuition is crucial to end-users since past research has shown that statically presenting only a few instances may lead to misrepresentation of the model limitations~\cite{kim2016examples} or the data that the model uses~\cite{vellido2012making}.
%The tool enables users to filter and then query the hidden states to find similar activations across the entire training set and overlays secondary properties on the results. 
Another desired outcome of interactive observational tools is the testing of hypotheses about local patterns that a model may have learned. For example, we developed Seq2seq-Vis~\cite{strobelt2019s}, which allows users to change inputs and constrain the inference of a translation model in order to pinpoint what part of the model is responsible for an error. This interaction is a type of non-collaborative shaping where the interactions are limited to the forward direction with the intended goal of understanding the decisions of the model.
Similar debugging-focused approaches~\cite{zhang2019manifold, krause2017workflow} only visualize the necessary parts of a model to find and explain errors, instead of giving in-depth insights into hidden model states.

%Due to the probabilistic nature of deep learning, defining quantitative measures for the outcomes is a challenging and unsolved problem. In our past work, we found it most helpful to run long-term studies and gather qualitative feedback instead. This enabled us to iteratively refine tools until the majority of the feedback confirmed the usefulness of the tool.

%\paragraph{Action} Instance-based passive observation techniques can be extended into interactive observations by including an interface for entering information or by precomputing the outcomes over a large set of data and providing a selection interface. For example, we developed LSTMVis to investigate whether active hidden states in a recurrent neural network are indicators of learned patterns~\cite{strobelt2018lstmvis}. While the tool does not allow evaluation of user-input, we precompute the states over the entire training set and enable search so that users can explore freely.

\subsection{Interactive Collaboration}

We characterize collaboration in interactive interfaces for deep learning models through the ability of end-users or trainers with domain knowledge to present feedback to the model (Figure~\ref{fig:intmodel}(c)) with the goal to shape the model or its decisions. We call these \emph{backward} interactions. 
Since each interaction direction needs to call a different shaping process within the model, the interface and model in interactive collaboration tools require a tight coupling. Only co-designing the model, visualizations, and interactions can achieve this tight integration. 

\paragraph{Model-Shaping}

On the model-level, the feedback is expressed as user-provided labels which can be used to change the model parameters in an active learning setting~\cite{holzinger2016interactive,cohn1996active,jiang2019recent}. 
In the forward direction, model performance can be visualized, and new samples for the labeling process can be selected \cite{holzinger2016interactive}. The model parameters can be updated through backward interactions~\cite{smilkov2017direct, kulesza2010explanatory}. 

\paragraph{Decision-Shaping} 

Interactive collaboration for decision-shaping requires an interface in which the end-user can guide the model-internal reasoning process to generate a different output than the model would have reached on its own. Since interactive collaborative interfaces also retain the ability for forward interactions, the intervention enables to an interplay between suggestions by the model and feedback by the user. Our proposed approach to developing CSI methods, which we describe in Section~\ref{sec:latvar}, presents one way to design such applications.\looseness=-1

% \paragraph{}
During forward interactions, the visualization shows what the model-internal reasoning process looks like for a specific input. During backward interactions, the end-user can modify the output and observe how the model-internal reasoning process would have looked like to arrive at that specific output. 
The incorporation of these feedback mechanisms into visual analytics tools requires three essential components. First, the model needs to expose an interpretable hook along its internal reasoning chain, which should be transparent in derivation and understandable for non-experts~\cite{lipton2016mythos}. Second, this interpretable hook needs to correspond to the mental model of the end-user. 
Most importantly, a collaborative tool needs to enable efficient interactions with the visual metaphor of the hook through semantic interaction~\cite{endert2012semantic}.  
%In this paper, we focus on describing interpretable hooks as latent variables within a model, a process that we describe in Section~\ref{sec:latvar}

%In addition, one can imagine the use of collaborative models as an extension of the user in a mixed-initiative user-inferface~\cite{horvitz1999principles} in use-cases where fully autonomous systems cannot be applied, for example in healthcare. Here, an associated measure is how the interface affects the efficiency and efficacy of the user compared to not using the tool.

The interpretable hooks of a model can act as explanations for rules of behavior that models learn. These explanations have been shown to improve model personalization~\cite{bostandjiev2012tasteweights} and explainability~\cite{caruana2015intelligible,kulesza2015principles}. Conversely, failing to provide explanations can inhibit the development of mental models in end-users~\cite{lim2009and}. However, explanations should also not overwhelm an end-user, and many previous approaches thus choose to select less complex models~\cite{lacave2002review} or aim to reduce the feature space of a trained model~\cite{craven1997using, yang1997comparative}.%\looseness=-1

CSI systems have models and end-users collaborate on the same output. This contrasts with previous work that often treats the model as a complementary assistant, for example recommending citations for a writer~\cite{babaian2002writer}.
Moreover, CSI argues for a design approach to collaborative interfaces where the user retains agency over the exposed parts of the model's reasoning process. 
Even in related approaches where the model and user both generate content, the users either do not have control over the model suggestions~\cite{guzdial2017general} or the model is replaced by uncontrollable crowdworkers~\cite{bernstein2010soylent}.
While previous work on interactive phrase-based machine translation showed promising results towards the goal of collaborative interfaces, the same techniques are not possible with deep learning-based approaches~\cite{green2014human}.
This lack of previous work (see Table~\ref{tab:cate}) can in part be attributed to the progress of deep learning methods, which have only recently reached the performance levels necessary for CSI-style interfaces. 

\section{Rearchitecting models to enable collaborative semantic inference}
\label{sec:latvar}

\begin{figure}[t!]
    \centering
    \includegraphics{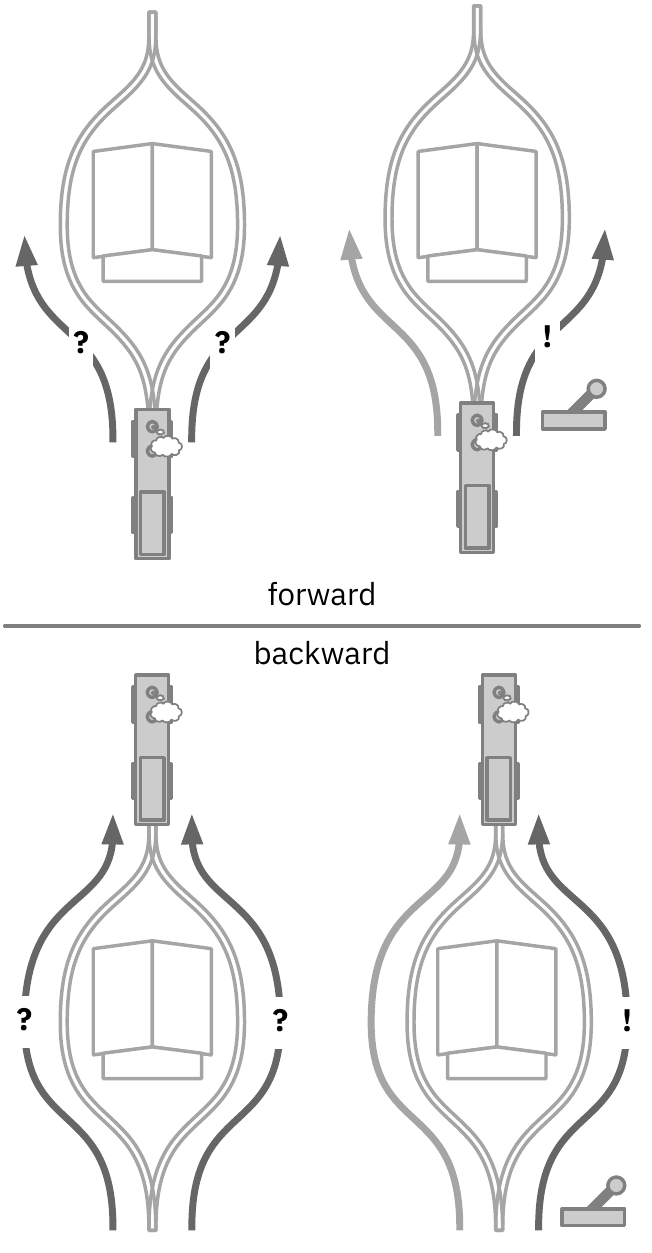}
    \caption{The advantage of latent variables is the transparent reasoning process. CSI requires model reasoning, illustrated by a lever that dictates which turn a train is taking. In the forward inference direction, we know in both cases where the train will end up, but only the lever allows us to know what path it will take. Similarly, in the backward inference direction, we know where the train originated, but only the lever shows the track it took to get there.}
    \label{fig:latvar}
\end{figure}

Interactive collaboration requires interpretable model hooks that enable semantic collaboration. From the machine learning side, these hooks can be implemented as discrete latent variables. During the prediction, or \emph{inference}, process, the variables take on explicit and understandable values. To illustrate this process, consider a hook resembling a lever that directs a train towards a left or right track, as shown in Figure~\ref{fig:latvar}. A model is predicting where a train will end up. Without the hook, the model can predict the end position accurately, but it is not clear how it will get there (top-left). Similarly, once the train reaches the top of the tracks, the model cannot explain how it got there (bottom-left). With the hook, however, the prediction explicitly exposes the lever decision. That means that it is predictable to the user whether the train will take the left or right branch (top-right). Once the train is at the top of the intersection, the user can look at the position of the lever to find the path the train has taken (bottom-right).

%In previous work, we have studied debugging tools for sequence models in order to examine predictions and conduct what-if analyses~\cite{strobelt2019s}. We found that the model architecture is a limiting factor. While we could change any intermediate representations and study outputs, small changes would often lead to major consequences. Furthermore, current systems only allow for \emph{forward} what-if analysis, whereas \emph{backward} inference, i.e., ``what internal decision would have led to a different output,'' is impossible.
%We propose using model "hooks" to target these issues by exposing internal model decisions. 

To formally describe hooks, we consider the problem of generating a sequence of words $y_1, \ldots, y_T$ that is conditioned on an input $x$ using a deep model. In the exemplary problem of document summarization, $x$ represents a long document and $y$ the summary. 
A forward-only deep sequence model defines a conditional distribution to predict one word at a time, $p(y_{t+1}\ |\ y_{1:t}, x)$ while considering all previously generated words. 
In the collaborative approach, the deep sequence model is extended to expose intermediate terms as latent variables $z$. In the summarization example, this could be the decision what words in an input are considered important enough to be included in a summary. 
The architect defines $p(y_{t+1}\ |\ y_{1:t}, x)= \sum_{z} p(y_{t+1}\ |\  y_{1:t}, x, z) \times p(z\ |\ y_{1:t}, x)$ for the latent variable $z$. That means that the model considers all possible values of $z$ to make a prediction. In the train-lever example, that means we can run the inference process to compute the best lever position by looking at which of the two positions would lead to a better final position as judged by the model. 

This decision splits the black-box into multiple parts: a \emph{prediction network}, $p(y_{t+1}\ |\  y_{1:t}, x, z)$, that predicts the next word, and a \emph{hook network}, $p(z\ |\ y_{1:t}, x)$, that predicts the value of the latent variables $z$. 
Because this model is probabilistic, we can also perform posterior or \emph{backward inference}. This gives $p(z\ |\ y_{1:T}, x)$, the distribution of $z$ after taking into account the entire output, i.e., the text summary. 

Another example of a collaborative interface could be for semantic image synthesis~\cite{chen2017photographic,wang2018high,Park2019semantic}. In this application, a user-defined input $x$ describes high-level features, for example, the location of grass and sky, and a neural network generates the corresponding image. In this case, $y$ is an image and not a sequence of words. The current approaches do not allow iterative refinement through interaction with an image and are limited to changing the input and generating a completely new output. A collaborative approach to the same problem is GANPaint~\cite{bau2018gandissect}, which uses a hook network to associate parts of the latent space of an image-generating model with the semantic features and exposes a modification interface to the user.

%For example, Hughes et al. develop a model for the classification of the correct antidepressant for a patient that constrains a set of latent variables to act as the explanation~\cite{hughes2018semi}. A visual interface for this model could enable the interaction with groups of these variables as an assistant to the doctor. 

In a real-world example of a hook-network, Google Translate recently introduced a semi-collaborative approach to prevent gender-discrimination in translation systems. By treating the gender as a hook, they can present both possible options whenever the gender in a language is ambiguous, for example in Turkish. This approach allows a user to pick the translation they want.

The model hooks represent ways in which users can constrain and direct interpretable predictions in otherwise end-to-end black boxes. They are extensions of otherwise well-performing models to expose the latent variable, co-designed by experts in interaction design, visualization, and machine learning. 
They have to decide on the model hooks, desired interactions, and the associated visual encodings. While some guidelines have been developed for interaction design for machine learning~\cite{fails2003interactive,stumpf2009interacting,amershi2014power,yang2018machine}, they focus on designing machine learning methods where the model performs complementary tasks to the user, or where the user interacts with black-box models. In contrast, CSI enables the study of interaction design for models that approach the same task as the user. 
One important question that requires further study is how many hooks are actually useful to a user. As more latent variables are designed and incorporated into a model, the training process becomes increasingly more challenging and the model performance might degrade. Moreover, the increase in potential user interactions with additional hooks might overwhelm users. As a consequence, we focus on a model with a single hook throughout our use case and show how a single hook can already enable many powerful interactions. 

We now describe our co-design approach for a CSI system for document summarization and present the lessons we learned from our CSI co-design process in Section~\ref{sec:process}.

\begin{figure*}[htb]
    \centering
    \includegraphics[width=\textwidth]{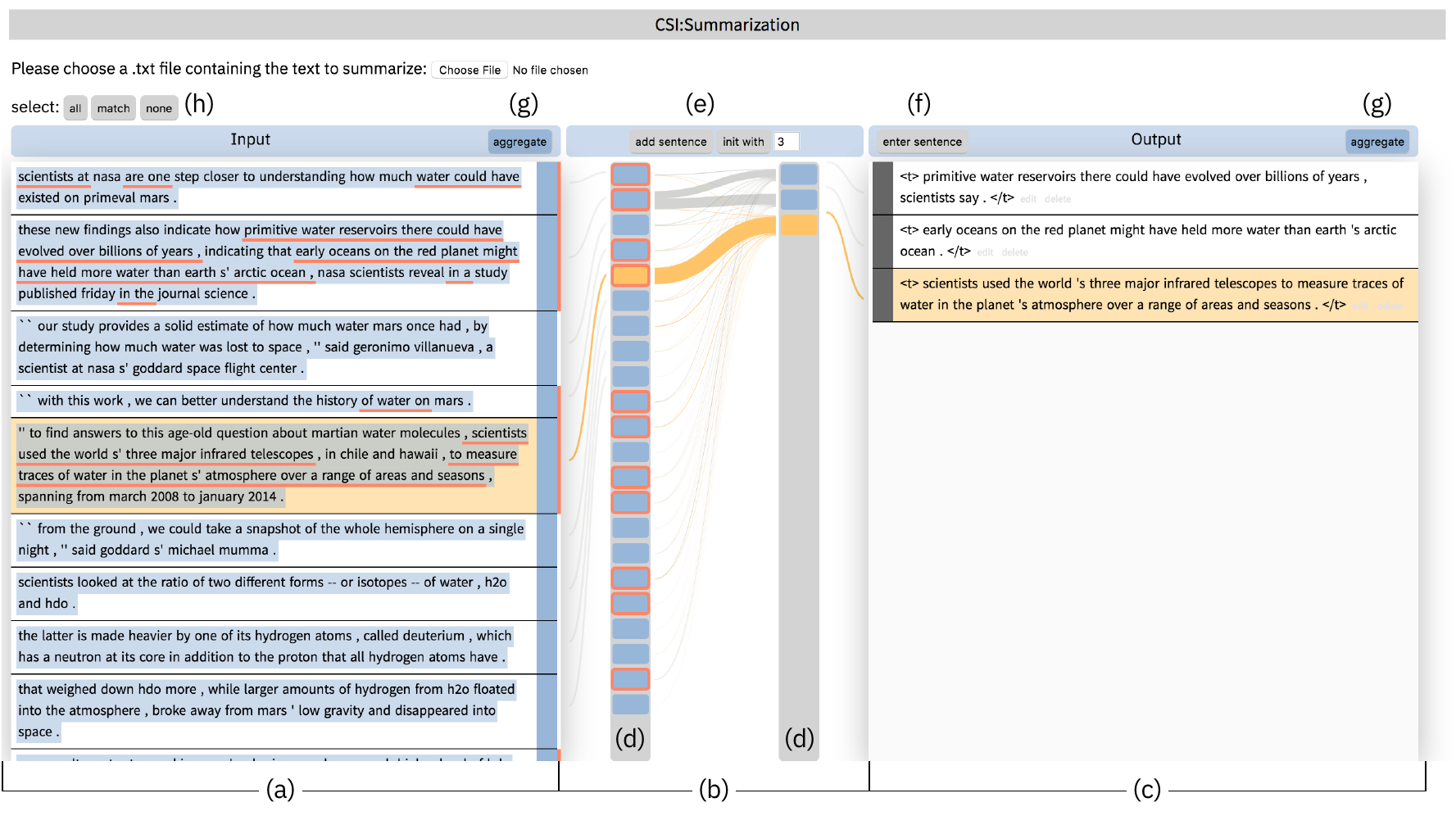}
    \caption{Overview of the CSI:Summarization visual interface. (a) shows the input and overlays the currently selected content selection (blue) and the current content of the summary (red). (b) shows a connection between the input and the current summary through the attention, which shows explicitly where words in the summary came from. (c) shows the current summary, (d) shows proxy elements for both input and output groups. This enables an overview of a document, even when the text does not fit on one page. (e) allows the user to request suggestions from the model, (f) enters the edit mode and adds a new sentence to the summary. (g) toggles whether the text should be aggregated into sentences. (h) provides quick selections for the content selection (blue) by being able to match the red highlights, or (de-)select everything.}
    \label{fig:sum-overview}
\end{figure*}

\section{Use Case: A Collaborative Summarization Model}
\label{sec:uc2}

% Structure
% \begin{itemize}
%  \item summarization problem (High level)
%     \item Current use of interfaces with post-editing -- forward path
%     \item introduce hook 
%     \item describe interactions enabled by it for forward and backward as main features in the interface
%     \item Introduce additional interface features
%     \item maybe: introduce some design challenges
% \end{itemize}

% Summarization Problem (high level)
%We design a CSI approach for the problem of document summarization. 
We demonstrate an application of the CSI framework to the use case of text summarization. 
Text summarization systems aim to generate summaries in natural language that concisely represent information in a longer text. 
This problem has been targeted by the deep learning community using an approach known as sequence-to-sequence models with attention~\cite{DBLP:conf/icml/XuBKCCSZB15}. These models have three main architectural components: (1) an encoder that reads the input and represents each input word as a vector, (2) a decoder that takes into consideration all previously generated words and aims to predict the next one, and (3) an attention mechanism that represents an alignment between the current decoding step and the inputs. In summarization, the attention can be loosely interpreted as the current focus of the generation and can be visualized for each generated word~\cite{strobelt2019s}.\looseness=-2

% Current use of interfaces / post editing
Imagine an end-user called Anna, who wants to use an interface powered by a summarization model. Current deep learning models act in a forward-only manner, as described above; therefore the design space is limited to an interactive observation interface. This interface allows Anna to paste an input text and have the model infer an output summary. If Anna does not like the output summary, she can edit the suggestion to her liking, but it is not possible to reuse the model to check her changes. Moreover, if the model produced a bad or wrong output, Anna would have to write the entire summary from scratch. 

%Neural-network generated summaries achieve high levels of fluency and coherence, but can create factual mistakes or select unimportant content to summarize. That means that generated summaries require heavy post-editing from Users, which is currently designed in a disconnected process. 

% Hook
Applying the CSI framework by extending a well-performing summarization model with the previously defined hooks can address these issues. By tying the user's interactions to understandable reactions within the model, we can achieve three previously not possible interactions. The collaborative interface (1) \textbf{guides the model} towards important content, (2) \textbf{enables the dialogue} between human-generated and machine-generated output, and (3) allows a user to \textbf{review the decisions} the model would have made to generate a specific output, i.e.,\ what parts of an input text the model chose to summarize.

These changes require designing a semantic model hook that can describe the content that a model considers for a summary. %take into account in the forward step. Then, the backward step can explain what content has been taken into account by predicting the position of the hook. 
We base the hook on a similar model we introduced in our previous work on document summarization~\cite{gehrmann2018bottom}. Commonly applied neural summarization models have the ability to either copy a word from the source document or generate a word. In contrast to users who write a summary, this approach does not have a planning phase where the model decides what content should be able to be copied from a more global perspective. 
We, therefore, introduce a hook for each word in a document that expresses whether the model should be able to copy the word. By exposing the hook within the interface, we can describe the semantic interaction as the user-decision what content of a document is relevant for its summary. A detailed description of the forward and backward models can be found in Appendix~\ref{app:sum}.\looseness=-1

%Internally, the user-input constrains the prior probability of the hook network, which means that the model is implicitly guided towards the selected content.

%However, the model is not overly constrained and can still generate a summary in any order or phrasing. 

%During the backward step, the model can use an entire document and its summary to predict what words and phrases in the input have been covered. This coverage takes into account even words that have been paraphrased in the output. %\hp{this paragraph is dense and a bit boring - loosing attention of the reader} 

% \sg{Hen: describe interactions enabled by it for forward and backward as main features in the interface}
%We co-designed the model modifications and the visual interface \textit{CSI:Summarization} (Fig.~\ref{fig:sum-overview}) to showcase a prototype for a collaborative semantic inference system \hp{repetitive - edit}. We note that this system incorporates almost all the necessary interactions \hp{necessary for what? not clear} to illustrate the complexity and depth of the design space \hp{how can we be sure that we explored the whole design space? seems implausible} and we will discuss potential simplifications later (Sec.~\ref{sec:vih_design}). \hp{simplify or delete previous paragraph}

We now describe how Anna generates a summary by using our CSI interface to collaborate semantically with the model.

\begin{figure*}
    \centering
    \includegraphics{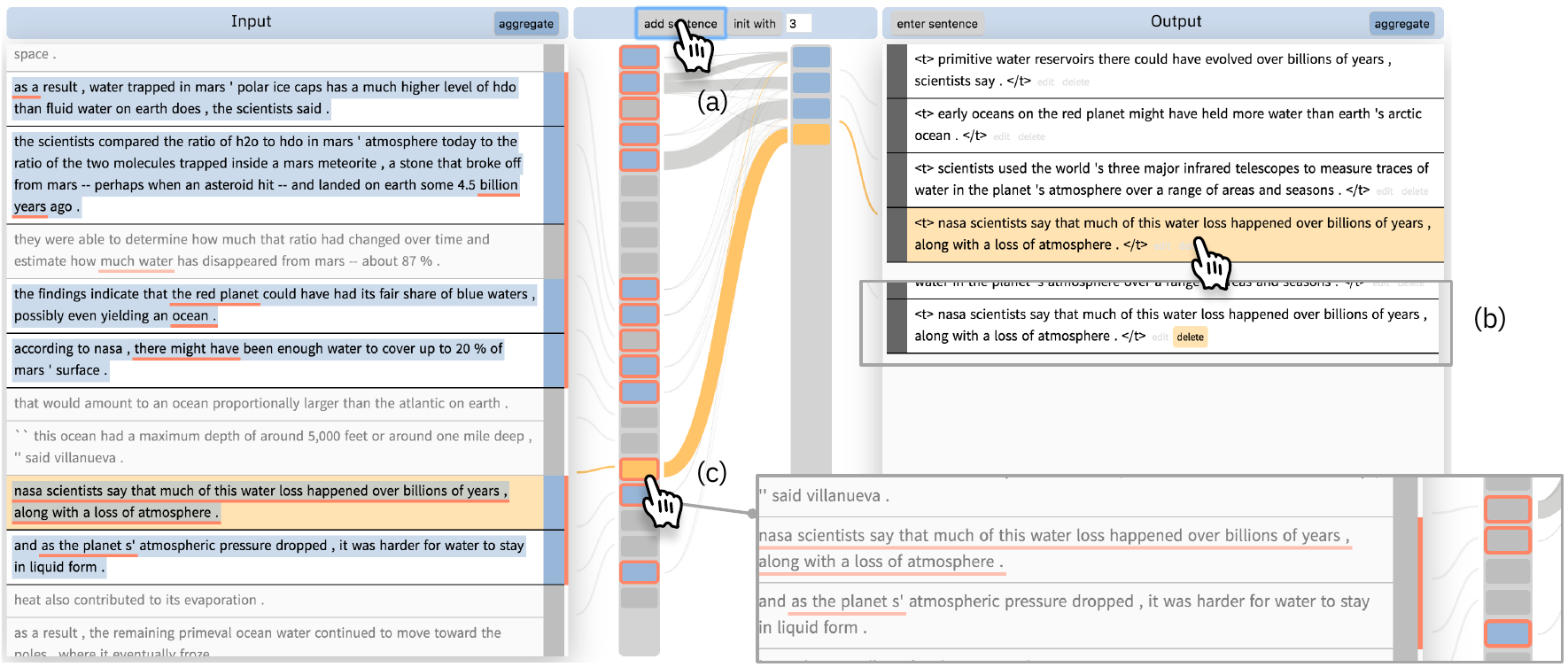}
    \caption{After selecting the content shown on the left, Anna requests the model to generate a fourth sentence (a). She does not like the suggestion and deletes it (b). To influence the model to generate about other topics in the input, she deselects the sentences that caused the suggestion (c). }
    \label{fig:uc_anna_01}
\end{figure*}

\begin{figure*}
    \centering
    \includegraphics{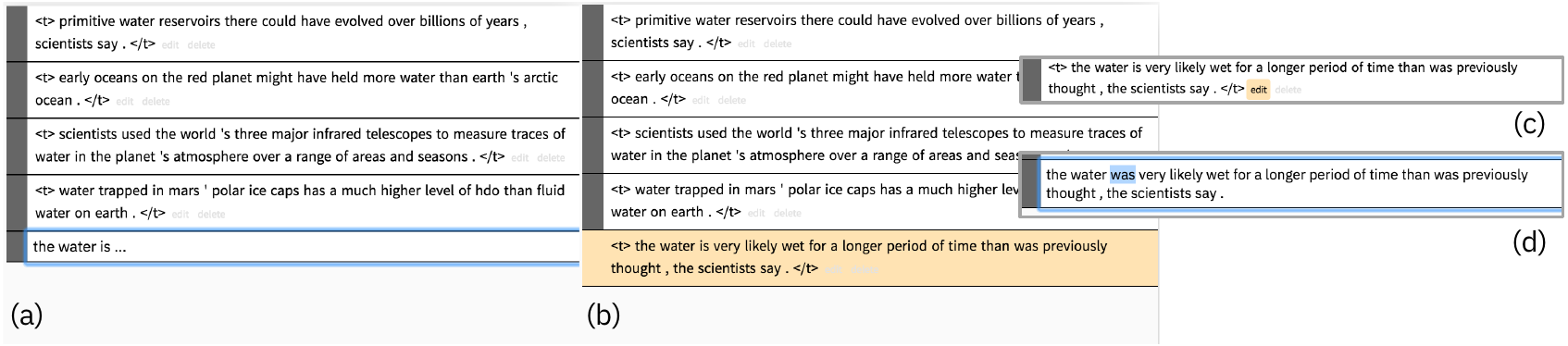}
    \caption{Anna wants to generate a sentence about the water on Mars and starts typing \emph{``The water is ...''} (a). This initiates the model to finish the sentence for her (b). To correct a minor mistake in the generated sentence, Anna activates the edit mode (c) and replaces the wrong word (d).}
    \label{fig:uc_anna_02}
\end{figure*}

\begin{figure*}
    \centering
    \includegraphics{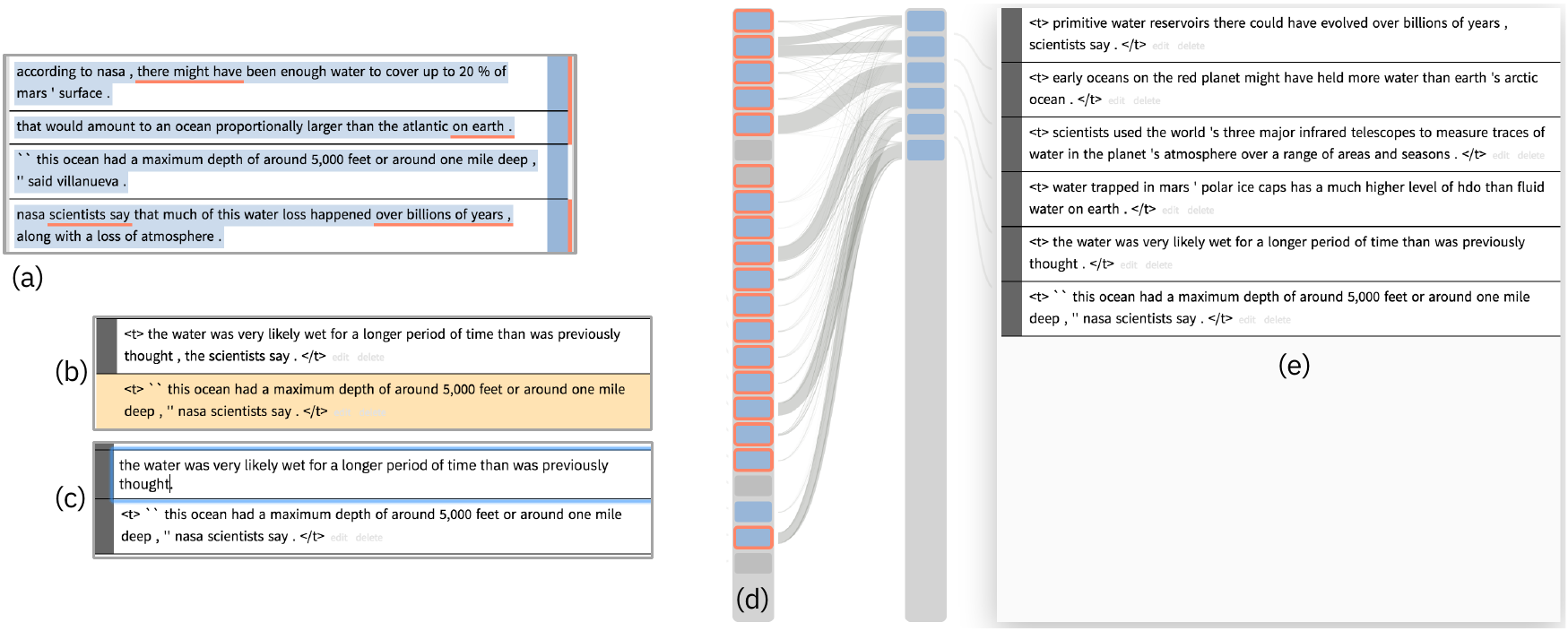}
    \caption{Anna selects a sentence that is currently not covered by the summary, indicated by the lack of a red border on the right (a). She generates a new sentence (b) and corrects the repeated phrase ``nasa scientists say'' (c). With the finished draft of her summary, she can now evaluate the coverage of the input document (d) and the final summary (e).}
    \label{fig:uc_anna_03}
\end{figure*}

\subsection{Collaborative Summarization: Anna's Story}

Anna intends to collaboratively write a summary of an article describing how scientists found water on Mars\footnote{The article can be found at \url{https://www.cnn.com/2015/03/06/us/mars-ocean-water-study}}. She begins by reading the article to assess what information is relevant and should be part of the summary. Then to begin the interaction, she selects the entire input text, shown by the blue highlights in Fig.~\ref{fig:sum-overview}a, letting the model know it is free to summarize any relevant part of the document. 

She starts the collaborative writing process by requesting that the model suggest three initial sentences (Fig.~\ref{fig:sum-overview}c). This triggers a forward inference of the model and a visual update that presents the suggestions to Anna. At the same time, the system computes a backward inference to show which part of the input has been summarized. Visually, the input text may be longer than the browser window, so we introduce a proxy element for each sentence in input and output (Fig.~\ref{fig:sum-overview}d). The covered words in the input are presented with a red underline and the proxies of the sentences with at least one covered word have a red border (Fig.~\ref{fig:sum-overview}a,b). The interface also visualizes the model ``attention,'' which shows which covered words were selected by the model at what step during the generation. The attention is visualized using grey ribbons that are aggregated across each sentence and connect the proxy elements. By hovering over one of the sentences or proxies, the interface highlights the relevant connections in yellow (Fig.~\ref{fig:sum-overview}b). 

Through these visual interactions with the output summary, Anna observes that the second input sentence (\emph{``scientists at nasa are one step closer to understanding how much water could have existed on primeval mars. these new findings also indicate how primitive water reservoirs there could have evolved over billions of years, indicating that early oceans on the red planet might have held more water than earth's arctic ocean, nasa scientists reveal [...]''}) splits into two different summary sentences (\emph{``primitive water reservoirs there could have evolved over billions of years, scientists say.''} and \emph{``early oceans on the red planet might have held more water than earth's arctic ocean.''}). It is common for summarization models to compress, merge, and split input sentences, but the user would not be aware of where the inputs for each summary sentence originate. By exposing the internal model decisions in a CSI interface, a user can immediately discover how the input connects to the output.\looseness=-1

From this suggested summary, Anna determines that the input focus of the system was correct, but that output text should elaborate on these sentences in more detail. She communicates this by first constraining the model to the 
currently focused region. She can match the content selection (blue highlights) with the result of the backward inference (red underlines) by clicking ``match'' in Fig.~\ref{fig:sum-overview}h.
 %This content is found during the backward step that the interface automatically triggered after presenting the suggestion. 
%She semantically interacts with the input mask and restricts it to all the sentences identified in the backward step for the current summary. 
With these constraints, she triggers the generation of an additional output sentence (\emph{``nasa scientists say that much of this water loss happened over billions of years, along with a loss of atmosphere.''}) (Fig.~\ref{fig:uc_anna_01}a). This suggestion is the result of a forward inference with her additional constraint on the model hook.\looseness=-1

Unfortunately, Anna is dissatisfied with the new sentence, so she removes it from the summary (Fig.~\ref{fig:uc_anna_01}b). To prevent the model from suggesting the same sentence again, she consults the backward inference and deselects the input sentence that had the highest influence on its generation by clicking on its proxy element (Fig.~\ref{fig:uc_anna_01}c). 
With this updated constraint, Anna generates another sentence (\emph{``water trapped in mars' polar ice caps has a much higher level of hdo than fluid water on earth.''}) that better captures her goals. (Fig.~\ref{fig:uc_anna_02}a). 

%Influencing the reasoning process of the model in this way is only possible by the semantic interaction with the hook that is enabled by the CSI model. 

Anna would next like to include more details to the summary, particularly about water found on Mars. The system allows her  to intervene in the output text directly. She starts writing ``the water is'' and then adds an ellipse (...) that triggers sentence completion by the model (Fig.~\ref{fig:uc_anna_02}b,c). The resulting sentence (\emph{``the water is very likely wet for a longer period of time than was previously thought, the scientists say.''}) is acceptable to her, but she now spots an error with the verb (\emph{``is''}) which should be in the past tense. She quickly corrects this output (Fig.~\ref{fig:uc_anna_02}d,e), which invokes a backward inference to the input document. By updating the red highlights in the input, the interface provides her with information about what content was selected was used to create this improved sentence. These interactions help her create a mental picture of the model behavior. 

Anna would finally like the model to help her generate a sentence about a region of the input that is currently not included in the summary. She selects a previously unused sentence in the input (\emph{``this ocean had a maximum depth of around 5,000 feet or around one mile deep, said villanueva.''}) (Fig.~\ref{fig:uc_anna_03}a), requests another forward inference, and approves of the resulting suggestion (\emph{``this ocean had a maximum depth of around 5,000 feet or around one mile deep, nasa scientists say.''}). However, she dislikes the repetition of ``scientist say'' (Fig.~\ref{fig:uc_anna_03}b). After entering the edit mode on the output side, she removes one of the repeated phrases (Fig.~\ref{fig:uc_anna_03}c). Upon leaving the edit mode, the interface automatically triggers another backward inference that updates which parts of the inputs are covered. Anna uses this information to evaluate how much of the document is covered by her summary. By looking at the computed coverage (Fig.~\ref{fig:uc_anna_03}d), she can observe which sentences are covered and analyze how many of the proxies have a red border. Her six sentences (Fig.~\ref{fig:uc_anna_03}e) summarize the original text pretty well.

\subsection{Visual and Interaction Design}
\label{sec:vih_design}

We designed the text summarization prototype (\textit{CSI: Summarization}) such that text occupies the majority of screen estate as the central carrier of information for the task. Two central panels (Fig.~\ref{fig:sum-overview}a,c) represent input text and output text. Each text box represents words that are aggregated into sentences. Text highlights in the input show information about the model hooks and relations between input and output. %The blue highlights at the word and sentence level indicate the current selection that the next forward pass of the model should consider. The red underlines (Fig.~\ref{fig:sum-overview}a) and borders indicate what inputs the model considered if it had generated the current output text, i.e., the output of the backward pass. Yellow highlights while hovering show connections between an input and the corresponding output and vice versa. 
Neutral gray colors are used on the output side to clearly distinguish them from the blue colors that represent selections on the input (Fig.~\ref{fig:sum-overview}c).\looseness=-1

\begin{figure}[t!]
    \centering
    \includegraphics{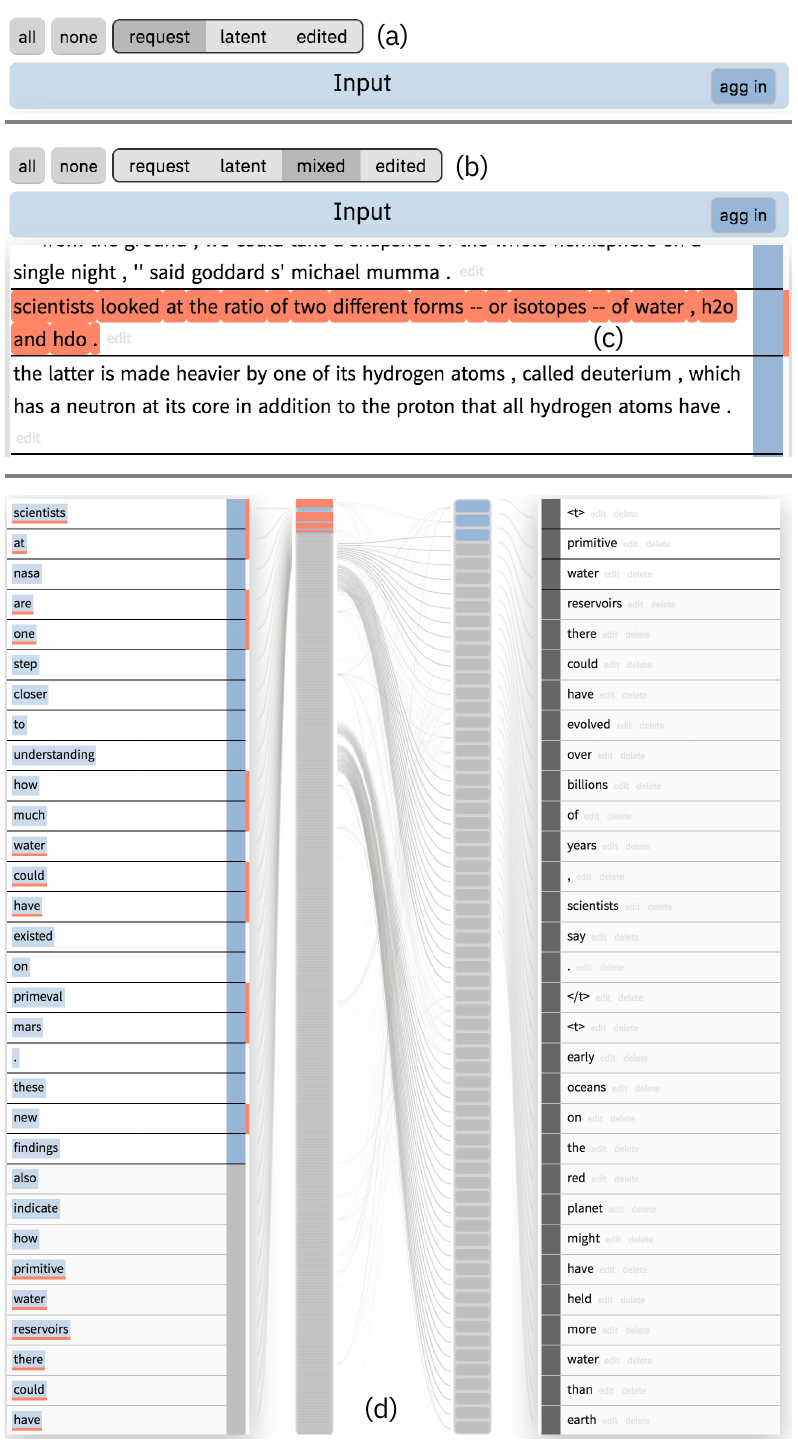}
    \caption{Visual design iterations for \textit{CSI:Summary}. (a) shows radio buttons for selecting a specific selection (content selection vs. backward model). In (b), we introduced a mixed mode that showed the user content selection in the final blue color and the result from the backward model with a red highlight (c). Underlines later replaced the dominant highlight. (d) shows an example of the complexity of the attention graph without any aggregation.}
    \label{fig:iteration}
\end{figure}

The input and output text are connected by a bi-partite graph that indicates model attention (Fig.~\ref{fig:sum-overview}b), which expands on previous work on visualizing and normalizing attention~\cite{strobelt2018debugging,linvisualizing}. %Similar to Seq2Seq-Vis~\cite{strobelt2019s}, the graph edges show what the model attended to during the forward inference. The attention is defined between all the words of input and output. 
Due to the length of source documents, displaying the entire graph is not feasible or informative. Therefore, we use two design elements to observe the full graph in a de-cluttered view: aggregations and proxies. First, we allow the aggregation of words into meaningful word groups, e.g., sentences, that can be dissolved on demand (Fig.~\ref{fig:iteration}d) if this level of detail is required. Aggregating words implicitly requires the aggregation of the attention which simplifies the graph. Secondly, we represent sentences by a vertical arrangement of boxes that are space-filling in height and which act as proxies for the full sentences. In that way, all sentences are always visible by their proxy, even when they are outside the display area. These proxies mirror selections and highlights of their related text boxes.\looseness=-1

\textit{CSI: Summarization} offers a range of user interactions. As a general principle, buttons trigger forward (left to right) actions (Fig.~\ref{fig:sum-overview}e) because a forward inference can change the output summary itself. Unintended changes to the output could confuse the user. To let users explicitly request updates instead of automatically intervening is inspired by similar mixed-initiative writing assistants~\cite{babaian2002writer}, where researchers found that this type of interaction is seen as least intrusive. %Moreover, the forward mode will change the summary, and unintended changes to the summary could confuse a user. 
All backward inferences are automatically triggered after exiting the edit mode by hitting enter. Since backward interactions do not change the content of the summary, the automatic invocation does not lead to accidentally overwriting important information. Users can define which sentences or words to consider for generating the next sentences by selecting them (blue color). Clicking on the bar on the right of an aggregation group selects or deselects the entire group. The same action is triggered by clicking on the proxy element of the corresponding sentence  (Fig.~\ref{fig:sum-overview}d). 

The interface additionally provides three selection templates  (Fig.~\ref{fig:sum-overview}h) for convenience: select all sentences, select no sentence, or select only those sentences that match the selection from the backward step (match red and blue). For the forward pass, the selection can be used to either initialize a new summary with a user-selected number of sentences (\texttt{init with}) or to add a sentence to the existing summary (\texttt{add sentence}) (Fig.~\ref{fig:sum-overview}e). On the output side, sentences can be deleted or edited by clicking the \texttt{edit} and \texttt{delete} buttons at the end of each sentence. %After finishing an edit on the output side, the backward model computes the corresponding input text that is being referred to, and the interface underlines the input accordingly. Similar to editing, the user can also create new sentences  (Fig.~\ref{fig:sum-overview}f). A new sentence can be completed by the model by typing an ellipse (...) end exiting edit mode. 

% We also considered an alternative interface in which a user can directly edit and add text without entering a separate mode. However, our decision to have a separate mode has three major advantages. Upon exiting the edit mode, we can provide (1) problem-specific automatic text formatting, and (2) look for the text finishing request \emph{...}. (3) The exiting of the edit mode is an explicit signal to initialize the backward step, which otherwise would need to be a separate button. 

\subsection{Design Iterations}

During the creation of the prototype, we explored multiple designs for model hooks, visualizations, interactions, and their integration. Overall, we found that CSI systems are more difficult to design because of the, sometimes competing, interactions between all these elements. We want to highlight one example for each of the elements. 

On the model side, our initial backward inference had good accuracy but did not reveal useful information within the interface. Only after re-allocating efforts to train a different model with a much higher performance did the backward model match the human intuition. Since model hooks complicate the design of the machine learning models, it also complicates their training process. Issues, such as the one described above, are only uncovered with the visual interface. 

On the visualization side, we explored multiple different designs for the input selections presented in the interface. The selections we currently show are (1) the selection that was used for the most recent forward step, (2) the selection that was returned from the most recent backward step, and (3) the sentences that the user is selecting for the next forward pass. In our first iteration, we aimed to show all of them in separate views (Fig.~\ref{fig:iteration}a), and additionally have a view that highlights their intersection (Fig.~\ref{fig:iteration}b,c). Internal tests revealed that only the combined view was useful to users to avoid having to switch forth and back. We, therefore, replaced the different views by the current more natural and coherent use of red and blue highlights within the same view. In conclusion, this design iteration provides a good example of the prevalent visualization challenge how to encode necessary information for the targeted user group. For CSI interfaces, the task is severe because of the major difference in abstraction between model internals and end-users intuitive understanding.

On the interaction side, we found that requesting the model to generate words without a constraint on the minimum or the maximum number of sentences often led to output that was unreasonable to users. The model architects on our team pointed out that the training data for summarization models rarely contains examples where the summary is longer than three sentences. Forcing a model to generate longer summaries than it was trained to generate led to the degradation in output quality. We also found that users had more control over the content of a summary if they iteratively built up a summary from a short initial suggestion instead of suggesting a lengthy summary and letting the user change it afterward. Our current modes combine these findings by designing the interaction after discussions with our model architects and visualization experts. 
The first interaction initially generates a small maximum number of sentences. This both leads to better model output but also lets users explore the output space more effectively. Similarly, adding one new sentence at a time by incorporating previous sentences as prefix context and allowing users to select the content enables users to quickly generate and review new content.

%\hp{can you add some text and a figure with design iterations for visualizations and interactions? currently, they seem to focus exclusively on the model}

% \sg{Hen: Introduce additional interface features}

% \sg{Hen: maybe: introduce some design challenges}

% \sg{Hen, visual design similarities, attention}

% Notes:
% - visual design considerations: 

% - visual interface shows input-model-output. user can explore how input objects is connected to output objects.

% - reduce visual clutter with edge bundling through proxies, emphasize objects in focus through highlighting/color coding.

% - keep overview by hierarchical aggregation of input and output objects. words to sentences. (maybe more important for second use case)

% - user can emphasize (but not enforce) objects/regions of interest in the input image, and output caption. - model takes this into account - collaboration.

% - further user actions: add/delete/edit object regions in the input, edit captioning in the output - visual interface shows most suitable matches to achieve this output in the input space.

% \subsection{Co-Design Iterations}

% \begin{itemize}
%     \item Aggregation for more efficient semantic interaction
%     \item Visual Codesign: backward model did not make sense, needed to update it. Vis helped identify when it was ``good enough''
%     \item Multiple modes -> one mode
%     \item Visual Metaphor: Blue/grey for selecting output vs input needed to be separated 
% \end{itemize}

\section{Towards a Co-Design Process for CSI Systems}
\label{sec:process}

% \begin{figure}
%     \centering
%     \includegraphics{figs/latvaria_process.pdf}
%     \caption{The CSI process model. The steps describe the core development process \hp{what does that mean? explain} and explicitly encode the co-design between machine learning and visualization experts.}
%     \label{fig:process}
% \end{figure}

During the implementation of the \textit{CSI:Summarization} prototype we developed an understanding of how an integrated design process for CSI systems could look like and also experienced its limitations. We discuss our insights as learned lessons.

\paragraph{Prioritize collaborative output.}
CSI systems enable joint production by model and end-user together. The resulting output should be the central element for developing visualization and interaction ideas. It is essential to evaluate if a \textit{CSI approach is beneficial} for the given task, e.g., a face recognition model used to unlock a cell phone does not benefit from the CSI approach as no shared output is produced. Since \textit{CSI methods are decision-shaping}, they require human oversight and interventions and are thus not suited for processing massive data. Moreover, since CSI interfaces are targeted at end-users, the visualization can be domain-specific but should abstract model-internals in an intuitive way. Finally, \textit{no side should dominate over the other} to allow model-human collaboration, which should be reflected in the visualization and interaction design -- e.g., allowing equal easy access to triggers for human input and model suggestions. 

\begin{figure}
    \centering
    \includegraphics[width=\columnwidth]{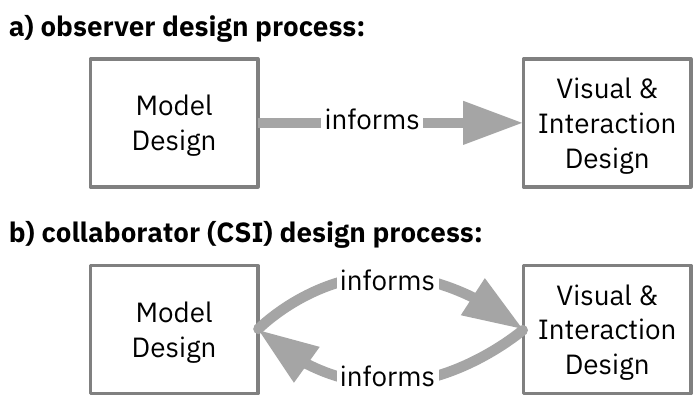}
    \caption{CSI interfaces require a design process that spans machine learning and visual interaction design. The feedback loop is used to produce interpretable semantic representations that inform the 
    reasoning procedure of the model while providing useful visual interaction. }
    \label{fig:loop}
\end{figure}

\paragraph{Co-design requires continuous evaluation.}
 In the development of the summarization system, there was a \textit{constant negotiation between visualization requirements and model capabilities}. This process led to iteration on the question: ``does this visual encoding help the end-user collaborating with the model?'' and ``which additional model behavior do we need to help to encode relevant information?''. Fig.~\ref{fig:loop} illustrates how this continuous co-design play forms a bilateral relationship between model design and visualization design. CSI systems, like most visual analytics systems, can help to \textit{reveal model problems} immediately. If, for example, a specific model hook is performing so poorly that it cannot facilitate the user's mental model, it will be immediately revealed. On the other hand, requirements for a \textit{visualization might over-constrain a model} such, that it breaks. E.g., creating a system for poem writing which should suggest lines of text that rhyme, even for human entered text, might over-constrain the model. CSI systems are about to find the middle ground between what would be ideal for a user interface vs. what would be possible with the underlying ML model.

\paragraph{CSI may be a worthwhile investment.}
Since CSI is centered around a single abstraction that should reflect the mental model that an end-user has of a problem, we pose that \emph{machine learning needs to learn about interactions}. There is currently a limited understanding of the space of easily trainable hooks and interaction strategies. Since machine learning techniques do not natively consider bidirectional interactions with end-users, the visualization, interaction design, and machine learning experts need to teach each other about desired interactions and the limits of deep models. Deep learning models with hooks thus lead to an increased development complexity for both machine learning and visualization experts. However, during the development of the summarization use case, we also experienced that \emph{CSI has a learning curve}. While CSI systems are individualized to a problem and thus one-of-a-kind systems, most of the techniques are transferable and the team can apply the insights gained from one CSI project to the next one. 
Moreover, the long-term benefits of the increased control over models can justify the additional development complexity, especially considering that many applications in industry are used for many years. As shown in our summarization example, the CSI methods can even lead to more structure and subsequent improvements in outcomes.

\section{Conclusions and Future Work}
\label{sec:conclusion}

In this paper, we introduced a framework for collaborative semantic inference, which describes a design process for collaborative interactions between deep learning models and end-users. Interfaces designed within this framework tightly couple the visual interface with model reasoning. We applied CSI to develop a collaborative system for document summarization that demonstrates that CSI systems can achieve powerful interactions within an interface powered by a neural model.\looseness=-1 
%We finally drew on the lessons we learned in the development of this and other related work and developed a focused process model that describes the development of CSI interfaces. 

While previous studies have shown that explainability methods can mediate in an agency-efficiency trade-off~\cite{lai2018human,yin2019understanding}, none of them demonstrate a way for users to retain agency while gaining the efficiency benefits of models interaction. We believe this is due to the difficulty of engaging the user in the prediction process of a black-box model. We address this problem by designing semantic interactions as part of the model itself. While CSI does not solve problems with biased data and models or the lack of interpretability of models, it aims to expose important model decisions and facilitate collaboration between an end-user and the model to take these decisions. Further developments in interpretability research could be used in conjunction with CSI for a better overall model understanding.  

The CSI framework significantly expands the interaction design space over conventional interaction strategies for many deep learning models. CSI-style approaches have the potential for application especially in scientific or safety-critical applications where explainable AI may become mandatory. Moreover, since CSI treats the model as a team member, another area of particular interest are creative applications, where models can assist users in creating stories~\cite{fan2018hierarchical}, chord progressions~\cite{huang2019music}, or even recipes~\cite{kiddon2016globally}. Future work might also investigate how similar principles could support cases where more than a single end-user and a single model aim to collaborate.\looseness=-1 

% A major future challenge towards progress on collaborative models is that current off-the-shelf deep learning models do not support the semantic inference necessary for CSI interfaces. 
% It is also crucial to note that not every problem benefits from CSI, for example all feature-driven applications such as architecture search and hyperparameter tuning. 
% While CSI models can expose a number of interpretable reasoning steps, they still suffer from similar black-box problems as other neural networks, since the rest of the model remains non-understandable. 
%While black-box explanations in interactive observation tools can yield efficiency increases while improving trust of humans in a machine learning model~\cite{lai2018human}, they can also lead to users trusting the information blindly~\cite{yin2019understanding}. 
%Thus, users retain agency over previously black-box approaches while simultaneously benefiting from the improved efficiency of deep learning models. 

Finally, it is crucial to develop ways to systematically evaluate collaborative interfaces and to investigate the implications of designing algorithmic interactions with humans~\cite{wilks2010close,williams2018toward}. 
While an interpretability-first approach could assist in highlighting fairness and bias issues in data or models~\cite{hughes2018semi, holstein2018improving}, it could also introduce unwanted biases by guiding the user towards what the model has learned~\cite{arnold2018sentiment}. It is thus insufficient to limit the evaluation of a system to measures of efficiency and accuracy. Future work needs to address these shortcomings by developing nuanced evaluation strategies that can detect undesired biases and their effect on end-users.

%By interacting with the visual interface, the users can avoid unethical output and similarly to prevent factual errors. 
%In first steps towards universally applied CSI models, Google translate recently introduced a semi-collaborative approach to prevent gender-discrimination in translation systems. By treating the gender as a hook, they can present both possible options whenever the gender in a language is ambiguous, for example in Turkish. This approach allows a user to pick the translation they meant, given the wider context and their intrinsic knowledge.

We provide a demo, the code, and the required models for CSI:Summarization at \url{www.c-s-i.ai}.% upon acceptance. 

%Furthermore, the use of these models is restricted by available human resources for the tasks they complete, whereas fully autonomous models require less supervision. In many use cases this is a benefit since it forces users of a model to double-check its outputs, but in others it might restrict the scalability. Finally, it is currently not well understood how these models can be formally evaluated without large-scale human studies. 

%We plan to further investigate CSI models in future work to help humans retain the agency over their automated processes and prevent unethical behavior of machine learning models. 

\section*{Acknowledgements}

We would like to thank Michael Behrisch for constructive discussions. This work was funded in
part by the NIH grant 5R01CA204585-02, an AWS Faculty Award and a Google Faculty Award.  
\appendix
%\section{Summarization Model}

%We illustrate one simple hook used with neural sequence models, which we call the \textit{copy hook}~\cite{vinyals2015neural,see2017get}.  In summarization, defined below, the goal is to generate a shorter $y$ from a document $x$ from a neural model.
%The copy hook modifies the summarization decoder to allow it to make a binary sub-decision $z$. At any point in time, it can choose to generate a word (in the normal fashion) or copy a word from the source
%document verbatim.  The hook network $p(z\ |\ y_{1:t}, x)$ makes the copying decision. The prediction network (M1) $p(y_{t+1}\ |\  y_{1:t}, x, z)$ takes that choice into account. If $z=0$ it generates a new word, if $z=1$ it copies a source word. Introducing this simple hook improves the accuracy of  summarization systems. Furthermore, backward inference facilitates powerful what-if case study analyses, e.g., Was a given word copied? Would a different word have been generated or copied instead?

\section{Details on the Summarization Model Hooks}
\label{app:sum}

Summarization models use source words $x_1, \ldots, x_n$ as input with the goal to generate a summary $y_1, \ldots, y_m$ with $m \ll n$.% A model represents as a function that aims to maximize the probability of generating the correct summary.
The most commonly used neural approach to document summarization is the sequence-to-sequence model~\cite{Bahdanau2015}, which encodes a source document and then generates one word at a time.
Sequence-to-sequence models incorporate an attention distribution $p(a_j|x,y_{1:j-1})$ for each decoding (generation) step $j$ over all the source words in $x$, which is calculated as part of the neural network. The attention can also be interpreted as the current focus of the model.
Since summaries often reuse words from the source document, current neural architectures incorporate a mechanism called copy-attention. This mechanism introduces a binary switch that enables the model to either generate a new word or copy a word from the source document during the generation~\cite{see2017get}. Similar to the standard attention, the copy-attention computes a distribution over all source tokens for each decoding step. 
The copy-attention (shown in Fig.~\ref{fig:sum-overview}e), is directly interpretable since a high value means that the model is actually copying a specific word from the input. 

While this approach yields a high performance on automatic metrics, the end-to-end approach with decisions per generated word is disconnected from human summarization approaches. 
There have been multiple studies that show that humans typically follow a two-step approach in which they first decide what content within a document is important, and then try to paraphrase it~\cite{jing1999decomposition}.
Following the goals of the CSI framework, we aimed to design a hook that reflects this two-step approach so that we can expose the planning stage and model decision what content is important to the end-user. Importantly, the end-user, in this case, may only have minimal or no knowledge of the underlying machine learning model, which means that every action by the end-user should be intuitive and directly connected to the hook. 

To enforce the human-like process within the model, we developed a CSI add-on for the high-performing model. We introduce binary tags for each of the source words, $t_1, \ldots, t_n$. The model is allowed to copy the $i$-th word only if $t_i$ is $1$\footnote{How to train the model is described in our previous work~\cite{gehrmann2018bottom}}.
The \emph{prediction network} that generates the summary $p(y_j|x, y_{1:j-1}, t)$ is thus prevented from copying the blanked out content. The additional \emph{hook network}, which predicts the tag probabilities $p(t|x)$, decides what content is important, which leads to significant improvements in fully automatic summarization. 
Since the model explicitly reasons over content-importance through the hook network, we can achieve semantic interactions by letting users define a prior on $p(t|x)$. When user deselects a sentence from the input, we set the prior $p(t)$ for all its words to 0, which means that the hook network can no longer identify the words as important which means that it is prevented from copying deselected words. 

The last step towards the fully integrated CSI:Summarization is the \emph{backward inference}, i.e. the identification of what content a summary actually used, or $p(t|x,y)$. The result of this is shown with red highlights in the interface in Fig.~\ref{fig:uc_anna_03}. The backwards model is a separate model we specifically developed for the interface. It uses a contextualized representation of words in both input and summary that represents them as vectors of size $d_{hid}$~\cite{devlin2018bert}. Given the representation for a word $x_i$, the model computes an attention over $y$, such that each summary word $y_k$ is assigned an attention weight $a_k$. We use these weights to derive a context for the word $x_i$ which we denote $c_i$, by computing

$$
    c_i = \sum_{k=1}^m a_k \times y_k.
$$

To arrive at a probability that $x_i$ was used in the output, we compute 

$$
   p(t_i|x,y) = \sigma(W_2\tanh(W_1[x_i, c_i] + b_1)+b_2),
$$

\noindent where $b_{1,2}$ are trainable bias terms and $W_1 \in \mathbb{R}^{d_{hid} \times 2d_{hid}}$ and $W_2 \in \mathbb{R}^{1 \times d_{hid}}$ trainable parameters. 
Since this model is independent of the forward model, it can analyze arbitrary summaries, even those that are user-written, as we show throughout our use case.

% COMMENT THIS FOR SUBMISSION
% \input{src/99-notes.tex}

%\acknowledgments{The authors wish to thank A, B, and C. This work was supported in part by a grant from XYZ (\# 12345-67890).}

%\bibliographystyle{abbrv}
\bibliographystyle{abbrv-doi}

\bibliography{template}

% \newpage
% \appendix
% \input{src/09-appendix.tex}
\end{document}